\documentclass[twocolumn,prb,showpacs,floatfix,superscriptaddress]{revtex4}
\usepackage{graphicx}
\usepackage{bm}
\newcommand{\be}{\begin{equation}}
\newcommand{\ee}{\end{equation}}
\newcommand{\bea}{\begin{eqnarray}}
\newcommand{\eea}{\end{eqnarray}}

\begin{document}

\title{Excitation spectra of the spin-$1/2$ triangular-lattice
Heisenberg antiferromagnet}

\author{Weihong Zheng}
\affiliation{School of Physics, University of New South Wales,
Sydney NSW 2052, Australia}
\author{John~O.~Fj{\ae}restad}
\affiliation{Department of Physics, University of Queensland,
Brisbane, Qld 4072, Australia}
\author{Rajiv R.~P.~Singh}
\affiliation{Department of Physics, University of California, Davis,
CA 95616, USA}
\author{Ross H. McKenzie}
\affiliation{Department of Physics, University of Queensland,
Brisbane, Qld 4072, Australia}
\author{Radu Coldea}
\affiliation{Department of Physics, University of Bristol, Bristol BS8 1TL,
United Kingdom}

\date{\today}


\begin{abstract}
We use series expansion methods to calculate the dispersion relation
of the one-magnon excitations for the spin-$1/2$ triangular-lattice
nearest-neighbor Heisenberg antiferromagnet above a three-sublattice
ordered ground state. Several striking features are observed
compared to the classical (large-$S$) spin-wave spectra. Whereas, at
low energies the dispersion is only weakly renormalized by quantum
fluctuations, significant anomalies are observed at high energies.
In particular, we find roton-like minima at special wave-vectors and
strong downward renormalization in large parts of the Brillouin
zone, leading to very flat or dispersionless modes. We present
detailed comparison of our calculated excitation energies in the
Brillouin zone with the spin-wave dispersion to order $1/S$
calculated recently by Starykh, Chubukov, and Abanov
[cond-mat/0608002]. We find many common features but also some
quantitative and qualitative differences. We show that at
temperatures as low as $0.1J$ the thermally excited rotons make a
significant contribution to the entropy. Consequently, unlike for
the square lattice model, a non-linear sigma model description of
the finite-temperature properties is only applicable at extremely
low temperatures.
\end{abstract}

\maketitle

\section{Introduction}
\label{intro}

Spin-$1/2$ Heisenberg antiferromagnets on frustrated lattices
constitute an important class of strongly correlated quantum
many-body systems. The interest in these models has been
particularly stimulated by the tantalizing possibility that the
interplay between quantum fluctuations and geometric frustration
might lead to a spin-liquid ground state and fractionalized (i.e.,
$S=1/2$) ``spinon''excitations. By a spin liquid we mean a state
which breaks neither translational nor spin rotational
symmetry. This exotic scenario originated with
the pioneering work by Anderson and Fazekas more than thirty years
ago,\cite{andfaz} where they suggested that a short-range resonating
valence bond (RVB) state might be the ground state of the
nearest-neighbor Heisenberg antiferromagnet on the triangular
lattice. More recently, considerable progress has been made in understanding
such states in terms of
field theory and quantum dimer models.\cite{slreviews} Whereas
 the
existence of such a ground state and of spin-half excitations is
well established in one dimension,\cite{giamarchi} it is
yet to be conclusively established theoretically in a realistic
two-dimensional Heisenberg model.\cite{moso}

Among the most important such models is the one
considered by Anderson and Fazekas, namely the Heisenberg
antiferromagnet on the triangular lattice with only nearest-neighbor
(isotropic) exchange interactions (hereafter just referred to as the
triangular-lattice model for brevity). However, over the past
decade numerical
studies\cite{bernu94,singhhuse,farnell,capriotti99} using
a variety of different techniques do not support
the suggestion in Ref. \onlinecite{andfaz} of a spin-liquid ground
state for this model. Instead, they provide evidence that the ground
state is qualitatively similar to the classical one, with
noncollinear magnetic N\'{e}el order with a three-sublattice structure
in which the average direction of neighboring spins differs by a 120
degree angle.

On the other hand, there are other theoretical results which suggest
that some properties of this model are indeed quite unusual. First,
a short-range RVB state is found to have excellent overlap with the
exact ground state for finite systems, much better
than for the square lattice.\cite{sindzingre,yunsor}
Second, variational calculations for RVB states, both with and
without long range order, give very close estimates for the ground
state energies.\cite{sindzingre,yunsor} Third, early
zero-temperature series expansion studies found some evidence that
this model may be close to a quantum critical point.\cite{singhhuse}
Fourth, one can make general arguments, based on the relevant
gauge theories\cite{gauge} that the quantum disordered phase
of a non-collinear magnet should have deconfined spinons,\cite{read}
although in the ordered phase the spinons are confined.\cite{chubukov2}
Finally, high-temperature series expansion studies\cite{elstner}
performed for temperatures down to $J/4$ ($J$ being the exchange
interaction) found no evidence for the ``renormalized classical''
behavior that would be expected from a semiclassical nonlinear sigma
model approach, if the ground state has long range
order.\cite{nlsmtr,css,css2} The actual behavior (summarized in some
detail in Sec. \ref{finiteT}) is rather striking and is in stark
contrast to the square-lattice model for which the ``renormalized
classical'' behavior appears very robust.\cite{chn89}

On the experimental side, there are currently no materials for which
it has been clearly established that their magnetic properties can
be described by  the $S=1/2$ isotropic triangular-lattice Heisenberg
antiferromagnet. In contrast, it has been clearly established that
the one dimensional and square lattice Heisenberg models with only
nearest-neighbour exchange give good descriptions of a number of
materials. Examples of the former  include KCuF$_3$\cite{tennant}
and Sr$_2$CuO$_3$,\cite{sandvik} and of the latter
Cu(DCOO)$_2$.4D$_2$O.\cite{mcmorrow}
For the triangular-lattice the most exciting prospect may be the
organic compound
$\kappa$-(BEDT-TTF)$_2$Cu$_2$(CN)$_3$,\cite{shimizu} which is
estimated from quantum chemistry calculations to have weak spatial
anisotropy\cite{komatsu}. Indeed comparisons of the thermodynamic
susceptibility with the high temperature expansions calculated for a
class of spatially anisotropic triangular-lattice models suggest
that the system may actually be very close to the isotropic
triangular-lattice Heisenberg model.\cite{zheng05hight} In Section
\ref{materials} we review the recent experimental evidence that this
material has a spin liquid ground state.

In this work we present series expansion calculations for the
triangular-lattice model. The primary focus is on the dispersion
relation of the magnon excitations above the 120-degree
spiral-ordered ground state. A brief description of some
of our results was
presented in an earlier communication.\cite{zheng06} In this paper
we discuss our results and the series expansion and extrapolation
methods in more detail. We also compare our results quantitatively
with very recent calculations of the spin-wave dispersion by Starykh
\textit{et al.}\cite{starykh} based on nonlinear spin-wave theory
which includes quantum corrections of order $1/S$ (to be called
SWT+1/S results) to the classical
large-$S$ or linear spin-wave theory (LSWT) results.

One of the most striking features of the spectrum is the local
minimum in the dispersion at the six wave vectors in the middle of
the faces of the edge of the Brillouin zone. Such a minimum is
absent in the spectrum calculated in linear spin wave theory. In
particular, along the edge of the Brillouin zone the semi-classical
dispersion is a maximum, rather than a minimum at this point. This
dip is also substantially larger than the shallow minima which
occurs in the square lattice model. Hence, this unique feature seems
to result from the interplay of quantum fluctuations and
frustration. We have called this feature a ``roton'' in analogy with
similar minima that occur in the excitation spectra of
 superfluid $^4$He\cite{feynman}
and the fractional quantum Hall effect.\cite{girvin} In those
cases, by using the single mode approximation for the dynamical
structure factor one can see how the roton is associated with
short range static correlations. Thus, an important issue is to
ascertain whether this is also the case for the minima we consider
here. Calculations of the static structure factor for the square
lattice do not show a minima at the relevant wavevector.\cite{zheng04}
We note that the roton we consider is quite distinct from the ``roton
minima'' for frustrated antiferromagnets that has been discussed by
Chandra, Coleman, and Larkin\cite{chandra}. The effect they discuss only
occurs for frustrated models which have large number of classically
degenerate ground states not related by global spin rotations. If we
apply their theory to the triangular lattice it does not predict such
a minima. Finally, we note that anomalous roton minima also appear
in the spectrum of the Heisenberg model with spatially anisotropic
exchange constants on the triangular lattice in the regime where the
magnetic order is collinear.\cite{zheng06} Such roton minima were
also found in a recent study of an easy-plane version of the same
model,\cite{alicea} where the elementary excitations
of the system are fermionic vortices in a dual field theory.
In that case, the roton is a vortex-anti-vortex excitation
making the ``roton'' nomenclature highly appropriate!

The spectra we have calculated
show substantial deviations from the LSWT results,
especially at high energies and for wavevectors
close to the crystallographic zone boundary, emphasizing the
importance of non-linear effects in the spin dynamics.
Several features of our calculated spectra are captured
 by the nonlinear spin-wave theory,\cite{starykh} but there are
also quantitative and qualitative differences. Both
calculations show a substantial downward renormalization of the classical
spectra. However, the highest excitation energies in the Brillouin Zone
are lowered with respect to LSWT by about $40\%$ in the
series calculations compared to about $25\%$ in SWT+1/S results.
Both calculations show substantial flat or nearly dispersionless spectra
in large parts of the Brillouin Zone. However,
the flat regions are much more pronounced in the series calculations
near the highest magnon energies, whereas they are more pronounced
at intermediate energies in SWT+1/S results. In the series calculations the
magnon density of states (DOS) has an extremely sharp peak near the
highest energies, whereas in the SWT+1/S calculations the largest peak in
DOS is at intermediate energies. The roton-like minima
at the mid-points of the crystallographic zone-boundary are much
more pronounced in the series calculations. They are much weaker
in the SWT+1/S results and are really part of the flat energy regions
contributing to the largest DOS peak in the latter calculations.

The SWT+1/S calculations are much closer to series expansion results
than LSWT and on this basis one can conclude that
the anomalous results obtained in series
expansions are perturbatively related to LSWT.
 In other words, a picture based on interacting magnons captures the
 single-magnon excitations, once the non-linearities
are taken into account. Indeed, we find that, if
we treat magnons as non-interacting Bosons and calculate their
entropy from the DOS obtained in the series calculations, we
get an entropy per spin of about $0.3$ at $T/J=0.3$, a value not far
from that calculated in high temperature series expansions.\cite{elstner}
Furthermore, we find that the low energy magnons give the
dominant contributions to the entropy only below $T/J\approx 0.1$.
This provides a natural explanation for why the non-linear
sigma model based description, which focusses only on the
low energy magnons, must fail above $T/J=0.1$.

There remains, however, an important open question with
regard to the spectra of this model. The question relates
to the nature of the multi-particle continuum above the
one-magnon states. In particular, how much spectral weight
lies in the multi-particle continuum,
and can it be described by an interacting-magnon picture, or
is it better thought of in terms of a pair of (possibly interacting) spinons?
This question is also related to the physical origin of the roton minima.
We note  that neutron scattering measurements
have observed a substantial multi-particle continuum in the
two-dimensional spin-1/2 antiferromagnet Cs$_2$CuCl$_4$\cite{coldea}
(which is related to the triangular lattice explored here, the main
difference being that Cs$_2$CuCl$_4$ has spatially anisotropic
exchange couplings). For that system  nonlinear spin-wave
theory\cite{veillette,dalidovich} could not account quantitatively
for the continuum lineshapes observed experimentally.

The plan of the paper is as follows. In Sec. \ref{model} we present
our model Hamiltonian. In Sec. \ref{expansions} we discuss the
series expansion methods used for studying zero-temperature
properties including the excitation spectra. Tables of various
series coefficients are also presented there. In Sec. \ref{extrapol}
we discuss series extrapolation techniques. After a short Sec.
\ref{gs} about ground state properties, we present results for the
magnon dispersion in Sec. \ref{tlm-spectra}, and compare them in
detail with nonlinear spin-wave theory. In Sec. \ref{thermo} we
consider how thermal excitation of the rotons affects thermodynamic
properties at much lower temperatures than might be expected, in
analogy with superfluid $^4$He. We show how this can explain the
absence of the renormalized classical behavior at finite
temperatures, well below $T=J/4$. In Sec. \ref{disc} we discuss a
possible interpretation of the roton in terms of confined
spinon-anti-spinon pairs and the relevance of our results to
experiments on $\kappa$-(BEDT-TTF)$_2$Cu$_2$(CN)$_3$. Finally, our
conclusions are given in Sec. \ref{concl}.

\section{Model}
\label{model}

\begin{figure}[h]
\begin{center}
  \includegraphics[width=6cm]{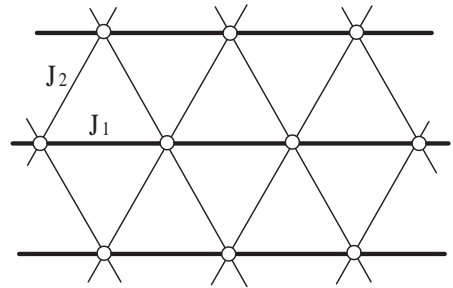}
  \caption{\label{trlattice}Exchange interactions in the Heisenberg model
  (\ref{HH}) on the triangular lattice. In this paper we focus on results
  for the case $J_1=J_2\equiv J$.}
\end{center}
\end{figure}

We consider an antiferromagnetic $S=1/2$ Heisenberg model on a
triangular lattice. More precisely we will analyze a 2-parameter
Hamiltonian of this type, given by
\begin{equation}
H = J_1\sum_{<in>}\bm{S}_i\cdot \bm{S}_n +
J_2\sum_{<ij>}\bm{S}_i\cdot\bm{S}_j.
\label{HH}
\end{equation}
Here the $\bm{S}_i$ are spin-$1/2$ operators. The first sum is over
nearest-neighbor sites connected by ``horizontal'' bonds (bold lines
in Fig. \ref{trlattice}) and exchange interaction $J_1$, the second
sum is over nearest-neighbor sites connected by ``diagonal'' bonds
(thin lines in Fig. \ref{trlattice}) with exchange interaction
$J_2$. As far as results are concerned, in this paper
our focus is on the isotropic model defined by $J_1=J_2\equiv J$,
but we find it convenient to distinguish between $J_1$ and $J_2$ for
the purpose of making our discussion of the series expansion (Sec.
\ref{expansions}) method more general.

\section{Series expansions}
\label{expansions}

In order to develop series expansions for the model in the ordered
phase, we assume that the spins order in the $xz$ plane, with an
angle $q$ between neighbors along $J_2$ bonds and an angle $2q$
along the $J_1$ bonds. The angle $q$ is considered as a variable;
the actual value of $q$ is that which minimizes the ground state
energy. We rotate all the spins so as to have a ferromagnetic ground
state, with the resulting Hamiltonian: \cite{zheng04,advphys,book}
\begin{equation}
H=H_1+J_1H_2+J_2H_3
\end{equation}
where
\be
H_1 = J_1\cos{(2q)}\sum_{<in>}S^z_iS^z_n +
J_2\cos{(q)}\sum_{<ij>}S^z_iS^z_j,
\label{eq_H1}
\ee
\be
H_2 = \sum_{<in>}S_i^yS_n^y+\cos{(2q)}S_i^xS_n^x +
\sin{(2q)}(S_i^zS_n^x-S_i^xS_n^z),
\label{eq_H2} \ee
\be
H_3 = \sum_{<ij>}S_i^yS_j^y+\cos{(q)}S_i^xS_j^x +
\sin{(q)}(S_i^zS_j^x-S_i^xS_j^z).
\label{eq_H3}
\ee
We introduce the Heisenberg-Ising model with Hamiltonian
\begin{equation}
H(\lambda)=H_0+\lambda V
\label{eq_final}
\end{equation}
where
\begin{equation}
H_0=H_1-t\sum_i (S_i^z-1/2),
\label{H0}
\end{equation}
\begin{equation}
V=J_1H_2+J_2H_3+t\sum_i (S_i^z-1/2).
\label{V}
\end{equation}
The last term of strength $t$ in both $H_0$ and $V$ is a local field
term, which can be included to improve convergence. At $\lambda=0$,
we have a ferromagnetic Ising model with two degenerate ground
states. At $\lambda=1$, we arrive at our Heisenberg Hamiltonian of
interest. We use linked-cluster methods to develop series expansion
in powers of $\lambda$ for ground state properties and the
magnon excitation spectra. The ground state properties are
calculated by a straightforward Rayleigh-Schr\"odinger perturbation
theory. However, the calculation of the magnon excitation
requires new innovations compared to a case of collinear order.
Since $S^z$ is not a conserved quantity here due to the last terms in Eqs.
(\ref{eq_H2}) and (\ref{eq_H3}), the one-magnon state and the
ground state  belong to the same sector. The linked-cluster
expansion with the traditional similarity transformation\cite{gel96}
fails, as it allows an excitation to annihilate from one site and reappear on
another far away, violating the assumptions for the cluster
expansion to hold. To get a successful linked-cluster expansion, one
needs to use the multi-block orthogonality transformation introduced
in Ref. \onlinecite{zhe01}. Indeed, we find that with proper orthogonalization the
linked-cluster property holds.

The series for ground state properties
have been computed to order $\lambda^{13}$, and the calculations
involve a list of  4\,140\,438 clusters, up to 13 sites. These
extend previous calculations\cite{singhhuse} by two terms, and
are given in Table \ref{tab_e0_M_ser}.
Since, we are working here with a model that has the full symmetry of the
triangular lattice, the series for the magnon excitation
spectra can be expressed as:
\begin{eqnarray}
\lefteqn{\hspace{-0.5cm}\Delta(k_x, k_y)/J =
\sum_{r=0}^{\infty} \lambda^r \sum_{m,n} c_{r,m,n} {\Big [}
\cos( \frac{m}{2} k_x) \cos(\frac{n \sqrt{3}}{2} k_y )} \nonumber \\
&+& \cos(k_y \sqrt{3} ( m + n)/4) \cos(k_x (m - 3 n)/4)  \nonumber \\
&+& \cos(k_y \sqrt{3} ( m - n)/4) \cos(k_x (m + 3 n)/4) {\Big ]}/3
\label{eq_mk_y1}
\end{eqnarray}
This series has been computed to order $\lambda^{9}$, and the
calculations involve a list of 38959 clusters, up to 10 sites. The
series coefficients $c_{r,m,n}$ for $t=1$ are given in Table
\ref{tab_mk_y1_t1}.

\section{Series extrapolations}
\label{extrapol}

In this section we discuss some details of the series extrapolation
methods used in our analysis. In order to get the most out of the
series expansions we have adopted a number of strategies.
The convergence of the series depends on the parameter $t$. This
parameter is varied to find a range where there is good convergence
over large parts of the Brillouin zone. However, the naive sum of the series
is never accurate at points where the spectra should be gapless. This is
true for any model and its reasons are explained below.

We have found it useful to also develop series for the ratio of our calculated
dispersion $\Delta(\bm{k})$ to the classical (large-S) dispersion $\Delta_{\rm{LSW}}(\bm{k})$
obtained from linear spin-wave theory. Following Ref. \onlinecite{merino},
$\Delta_{\rm{LSW}}(\bm{k})$ for arbitrary $\lambda$ and $t$ is given by
\be
\Delta_{\rm{LSW}} ( \bm{k} ) = 2 S \sqrt{(\lambda A+C) (\lambda B+C)}
\ee
where
\begin{eqnarray*}
A &=& J_1 \cos (k_x) + 2 J_2 \cos (k_x/2) \cos ( \sqrt{3} k_y/2), \\
B &=& J_1 \cos (k_x) \cos (2 q) \\
  &+& 2 J_2 \cos ( \frac{k_x}{2} ) \cos (
\frac{\sqrt{3}}{2} k_y ) \cos (q),  \\
C &=& 2 t ( 1 - \lambda ) - J_1 \cos (2 q) - 2 J_2 \cos (q).
\end{eqnarray*}
We can expand $\Delta_{\rm{LSW}}(\bf{k})$ in powers of $\lambda$,
and the ratio of our series expansion calculation $\Delta(\bf{k})$ to the series for
this linear spin-wave energy $\Delta_{\rm{LSW}}(\bf{k})$ will be called the ratio series
for the rest of the paper. The naive sum of this ratio series appears to converge
better because to get estimates for $\Delta(\bf{k})$ from it, we need to multiply the sum
by the classical energy $\Delta_{\rm{LSW}}(\bf{k})$ and this ensures that both
vanish at the same  $\bf{k}$-points.

We have also done a careful analysis of the series using series extrapolation
methods. By construction, the Hamiltonian
$H(\lambda)$ has an easy-axis spin-space anisotropy for $\lambda<1$, which
leads to a gap in the magnon dispersion. This anisotropy goes away in
the limit $\lambda\to 1$ when the Hamiltonian becomes SU(2)-invariant.
In this limit the gap must also go away as long as the ground state breaks
SU(2) symmetry. This closing of the gap is known to cause singularities
in the series. The singularities are generally weak away from ordering wavevectors
and gapless points, but are dominant near the ordering wavevector
where the gap typically closes in a power-law manner
in the variable $1-\lambda$.\cite{advphys,book}

We have used d-log Pad\'{e} approximants and integrated differential approximants
in our analysis. In general, these approximants represent
the function of interest $f$ in a variable $x$ by a solution to
a homogeneous or inhomogeneous differential equation,
usually of first or second order, of the form
\be
P_K (x) {d^2 f\over d x^2} + Q_L (x) {d f\over d x} + R_M (x) f + S_T (x) =0
\ee
where $P_K$, $Q_L$,  $R_M$, $S_T$ are polynomials of degree $K$, $L$, $M$,
$T$ respectively. The polynomials are obtained by matching the coefficients
in the power series expansion in $x$ for the above equation. They
are uniquely determined from the known expansion coefficients
of the function $f$ and can be obtained by solving a set of linear equations.
If $P_K$ and $S_T$ are set to zero, these approximants
correspond to the well known d-log Pad\'{e}
approximants, which can accurately represent power-law behavior.
Integrated differential approximants have the additional advantage
that they can handle additive analytic or non-analytic terms which
cause difficulties for d-log Pad\'{e} approximants. It is also
possible to bias the analysis to have singularities at predetermined
values of $x$ with or without predetermined power-law exponents.
Using such approximants, which enforce a certain type of
predetermined behavior on the function, is called biased analysis.
We refer the reader to Ref. \onlinecite{book} for further details.

\begin{figure}[!htb]
\begin{center}
  \includegraphics[width=8cm]{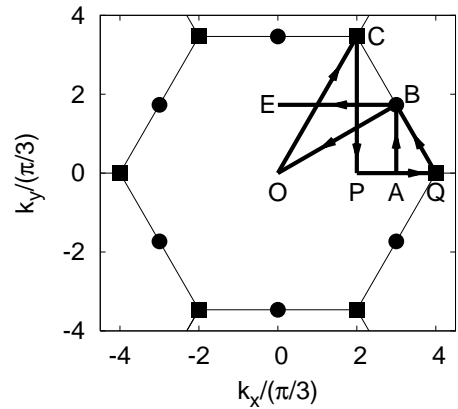}
  \caption{\label{trlattbz}
Reciprocal space of the triangular lattice including the hexagonal
first Brillouin zone. Squares denote ordering wavevectors, circles
denote wavevectors of the ``roton'' minima. The labeled points have
coordinates ${\rm O}=(0,0)$, ${\rm P}=(2\pi/3,0)$, ${\rm
A}=(\pi,0)$, ${\rm B}=(\pi,\pi/\sqrt{3})$, ${\rm
C}=(2\pi/3,2\pi/\sqrt{3})$, ${\rm Q}=(4\pi/3,0)$, and ${\rm
E}=(0,\pi/\sqrt{3})$. Also shown is the path {\rm ABOCPQBE} along
which the magnon dispersion is plotted in Figs.~\ref{dirn-plota} and
\ref{dirn-plotb}.}
\end{center}
\end{figure}

We found that the convergence near the ordering wavevector $Q$
(see Fig. \ref{trlattbz}), was particularly poor.
We know that we must have gapless spectra at $k=Q$
and $k=0$ as long as there is long-range
order in the system. Yet, most unbiased analysis gave a moderate gap
at $k=Q$.
The convergence is better near $k=0$, where unbiased analysis
is consistent with very small values of the gap.
This behavior near $Q$ may be some
evidence that long-wavelength correlations are not fully captured
by the available number of terms in the series.

\section{Ground state properties}
\label{gs}

In this section we briefly discuss results for two ground-state properties of the
triangular-lattice model: the ground state energy per site $E_0/N$ and the N\'{e}el
order parameter $M$ (i.e., the sublattice magnetization).

\begin{figure}[!htb]
\begin{center}
  \includegraphics[width=6cm]{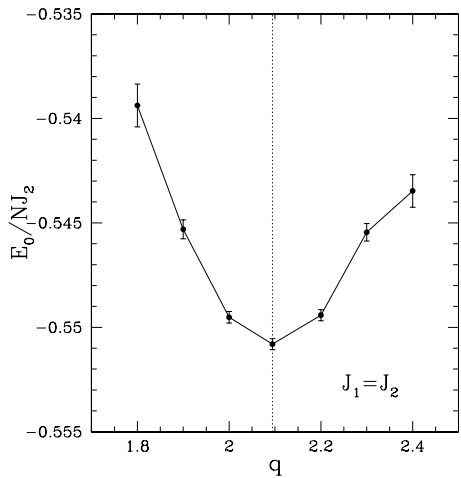}
\caption{\label{fig_e0}
The ground state energy per site $E_0/N$, as a function of the angle $q$
between nearest neighbor spins along $J_2$ bonds, for the triangular-lattice
model (i.e., $J_1=J_2\equiv J$).
The minimum energy is obtained when $q=2\pi/3$, the same as for the classical model.}
\end{center}
\end{figure}

In Fig. \ref{fig_e0} we show the extrapolated ground state energy as a function of
the angle $q$ between nearest neigbor spins along $J_2$ bonds. Clearly the ground state
energy is minimized when $q$ takes the classical value $2\pi/3$. The resulting value
for the ground state energy is $E_0/N = -0.5502(4)J$, which compares well with results
obtained from other methods (see Table \ref{encomparisons}).

The series for the order parameter $M$ is extrapolated assuming a square-root singularity
at $\lambda=1$, i.e., we extrapolate the series in the variable
$\delta=1-(1-\lambda)^{1/2}$ using integrated differential approximants. This leads
to the estimate $M=0.19(2)$. This estimate is known to be sensitive to the choice of the
power law.\cite{singhhuse}
Our value for $M$ shows good consistency with what is obtained from other methods (see
Table \ref{encomparisons}).

\begin{figure}[!htb]
\begin{center}
  \includegraphics[width=7cm]{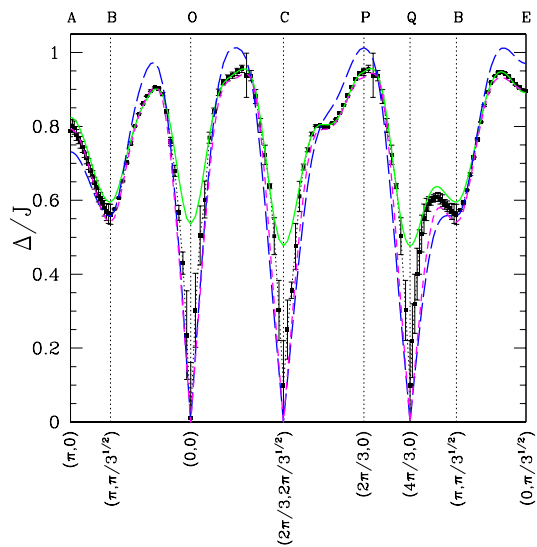}
  \caption{\label{dirn-plota}
(Color online) Calculated spectra along ABOCPQBE
of the Brillouin zone. Series extrapolation results (data points with error bars)
are plotted together with
naive sum of series with $t=2$ (green curve) and
naive sum of ratio series with $t=1$ (blue dashed) and $t=2$ (magenta dashed).
}
\end{center}
\end{figure}

\begin{figure}[!htb]
\begin{center}
\includegraphics[scale=0.35]{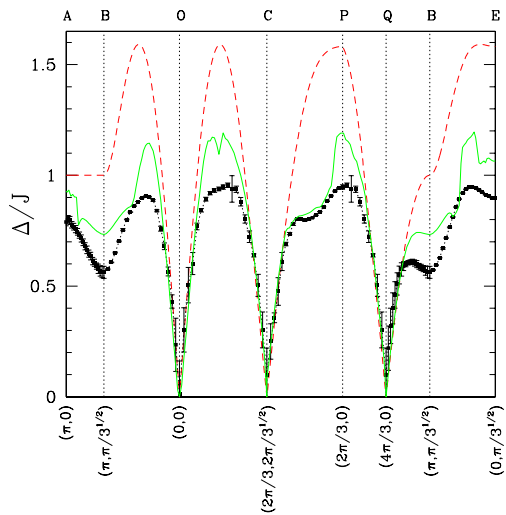}
\caption{\label{dirn-plotb} (Color online)
Magnon spectra along ABOCPQBE from series expansions compared with LSWT
(dashed red line) and SWT+1/S (green line).}
\end{center}
\end{figure}

\section{Excitation spectra}
\label{tlm-spectra}

The triangular-lattice Brillouin zone with selected wavevectors is
shown in Fig.~\ref{trlattbz}. In Fig.~\ref{dirn-plota} we plot our
most carefully extrapolated spectra along selected directions of the
Brillouin zone using integrated differential approximants with
appropriate biasing near the gapless points. The error bars are a
measure of the spread in the extrapolated values from different
approximants. Also shown in figure are the results from naive
summation of the series as well as naive summation of the ratio
series (see section IV) with two different $t$ values. In
Fig.~\ref{dirn-plotb} we plot the series expansion results together
with LSWT and SWT+1/S spectra.

The comparisons in Fig.~\ref{dirn-plota} show that the substantial
depression in the spin-wave energies obtained in the series expansions,
over large parts of the Brillouin zone,
is a very robust result that does not depend on extrapolations.
Roton-like minima at wavevector B is also a
very robust result already present in naive summation of the series.
On the other hand, series extrapolations are essential near gapless points O, C,
and Q. Even though the ratio series gives gapless excitations at these points,
it does not get the spin-wave velocity right.

\begin{figure}[!htb]
\begin{center}
\includegraphics[scale=0.35]{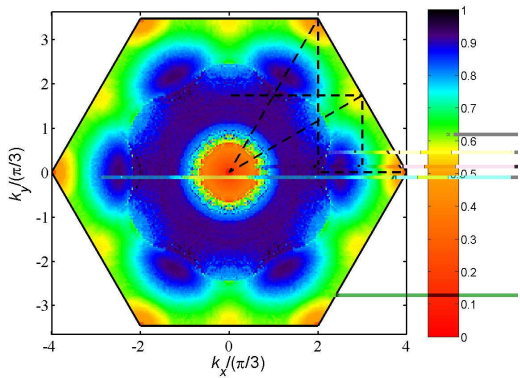}
\caption{\label{pro_series} (Color online) Projection plot showing the
magnon energies obtained from series expansions
in the triangular-lattice Brillouin zone.}
\end{center}
\end{figure}

\begin{figure}[!htb]
\begin{center}
\includegraphics[scale=0.35]{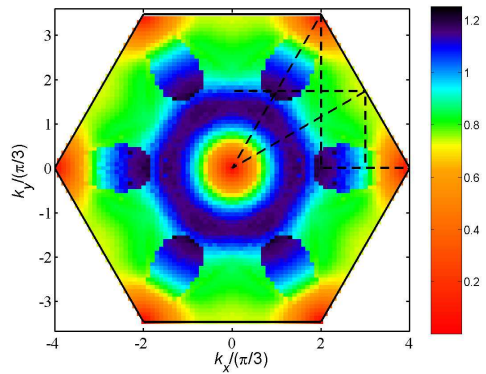}
\caption{\label{pro_swt+1/s} (Color online) Projection plot showing the
SWT+1/S magnon energies in the triangular-lattice Brillouin zone.}
\end{center}
\end{figure}

Since it is tedious to perform the full analysis of the spectra at all
points of the Brillouin zone (and not necessary for the higher
energy spectra), we have instead carried out a more restricted D-log Pad\'{e}
analysis over the whole zone.
In Fig.~\ref{pro_series} we show a two-dimensional projection plot of
the spectra in the full Brillouin Zone.
The color-code is adopted to
highlight the higher energy part of the spectra, where our results
should be most reliable, and minimize the variation at low energies
where this analysis is not reliable. In Fig.~\ref{pro_swt+1/s},
the corresponding two-dimensional projection plot for the SWT+1/S
calculations are shown.

\begin{figure}[!htb]
\begin{center}
\includegraphics[scale=0.5]{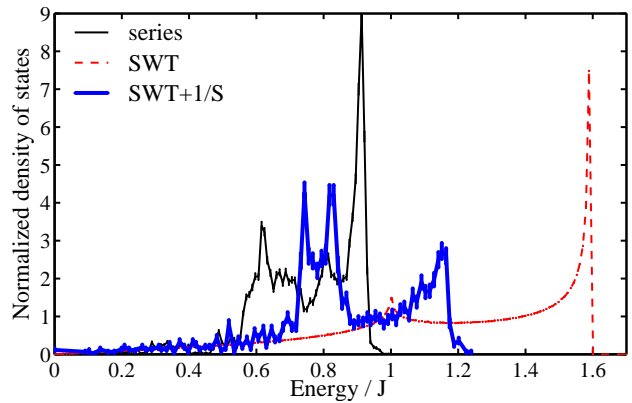}
\caption{\label{dosplot} (Color online) Plots of magnon density of states
for the series expansions, LSWT and SWT+1/S spectra.}
\end{center}
\end{figure}

In Fig.~\ref{dosplot}, we show the density of states (DOS) obtained
from series analysis, LSWT and SWT+1/S. In each case the integrated
density of states is normalized to unity.

From these plots, we make the following observations:

1. The SWT+1/S results share many common features with the series expansion
results. Over most of the Brillouin zone the SWT+1/S results fall in
between LSWT and series expansion results. They show that quantum fluctuations lead to
substantial downward renormalization of the higher energy magnon
spectra. This is in contrast to unfrustrated spin models, such as square-lattice or
linear chain models, where quantum fluctuations lead to increase in excitation energies.

2. There are quantitative differences in the downward renormalization.
The highest magnon energies are lowered with respect to LSWT by about
$40\%$ in the series results and by about $25\%$ in SWT+1/S results.

3. The agreement in the low energy spectra and the spin-wave
velocities is good when SWT+1/S results are compared to the biased
integrated differential approximant analysis of the series.

\begin{figure}[!htb]
\begin{center}
\includegraphics[scale=0.35]{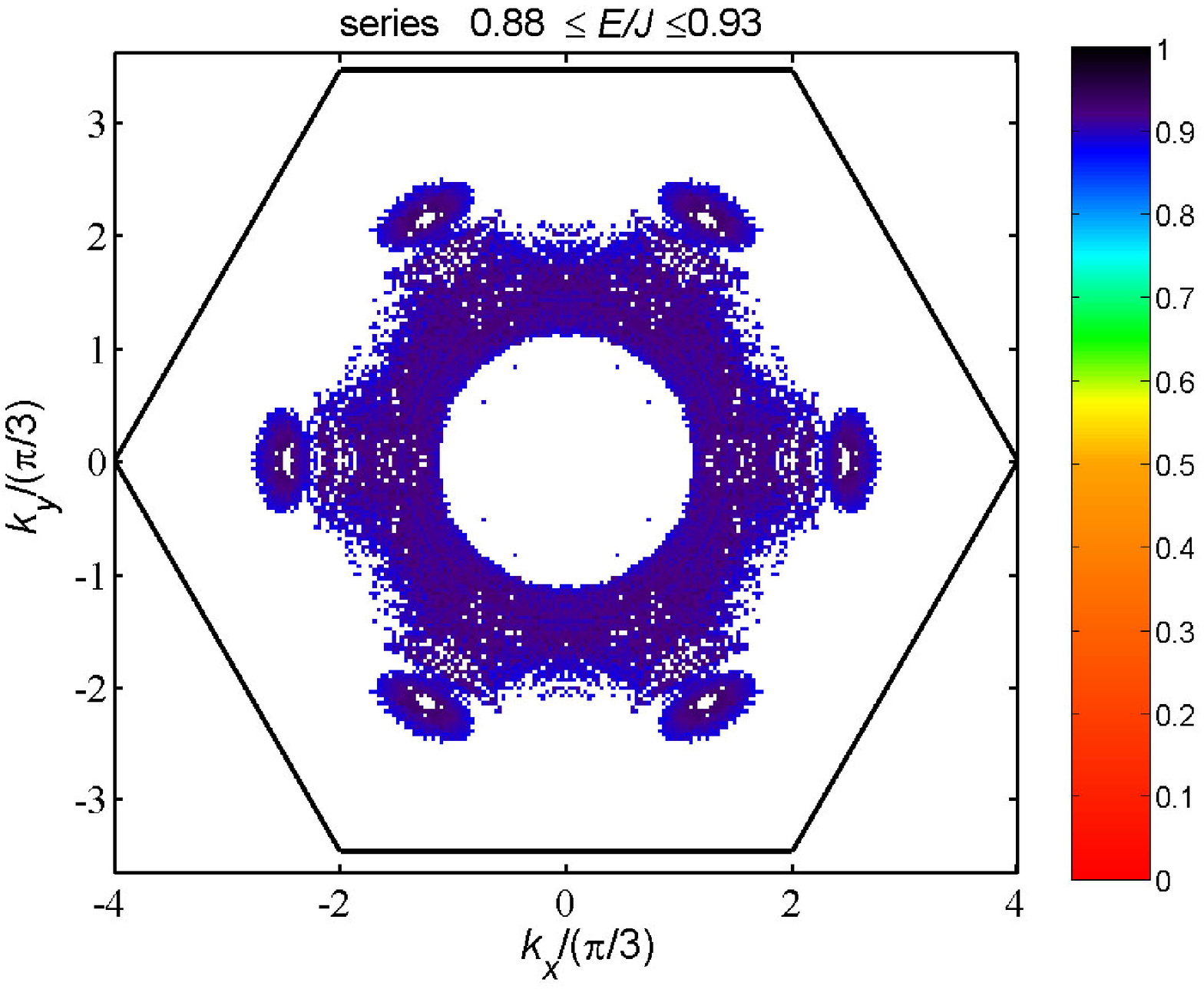}
\includegraphics[scale=0.35]{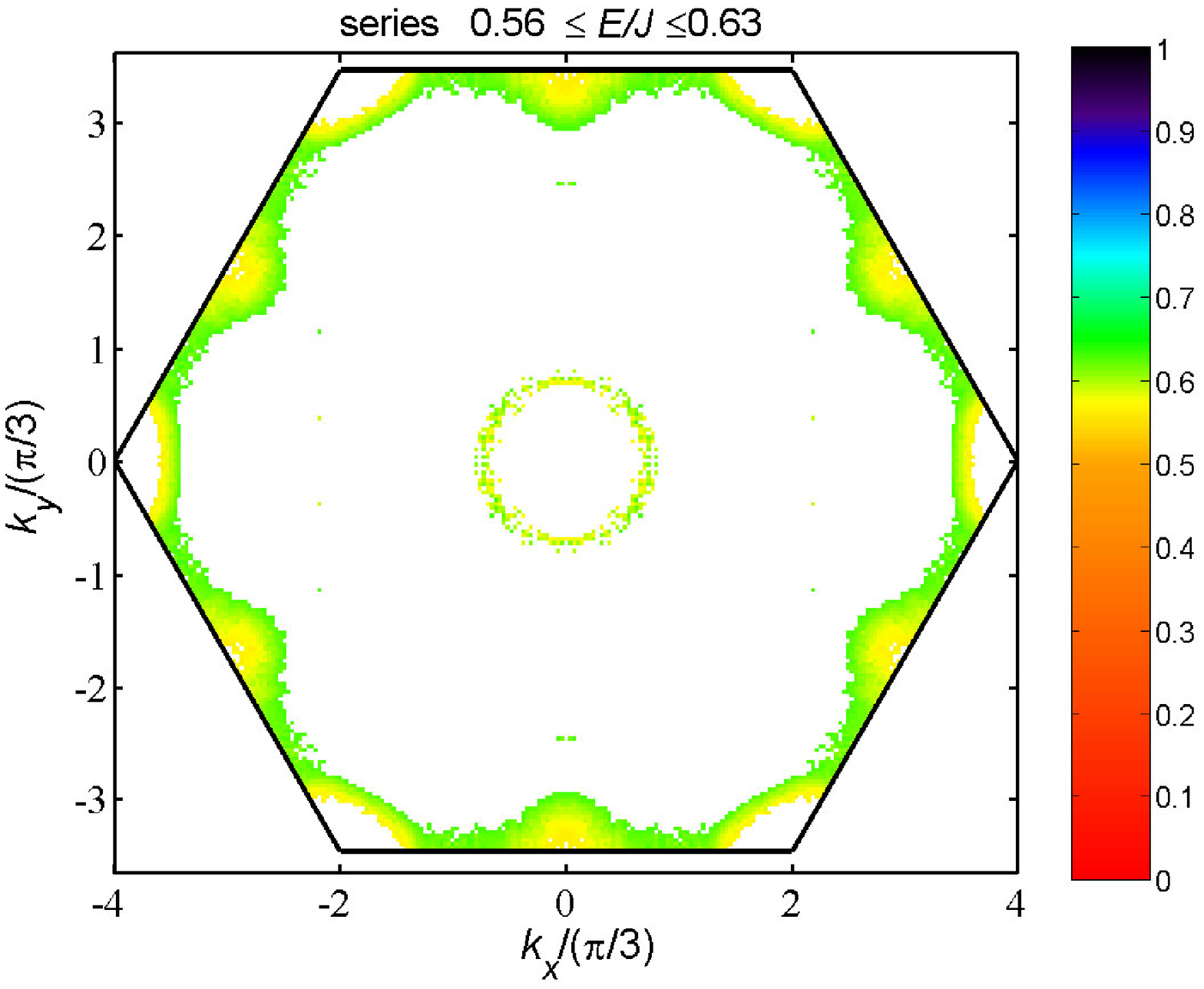}
\caption{\label{dos-peak-series} (Color online) Highlight of regions
in the Brillouin zone that contribute to the DOS peaks
in the series calculations.}
\end{center}
\end{figure}

\begin{figure}[!htb]
\begin{center}
\includegraphics[scale=0.35]{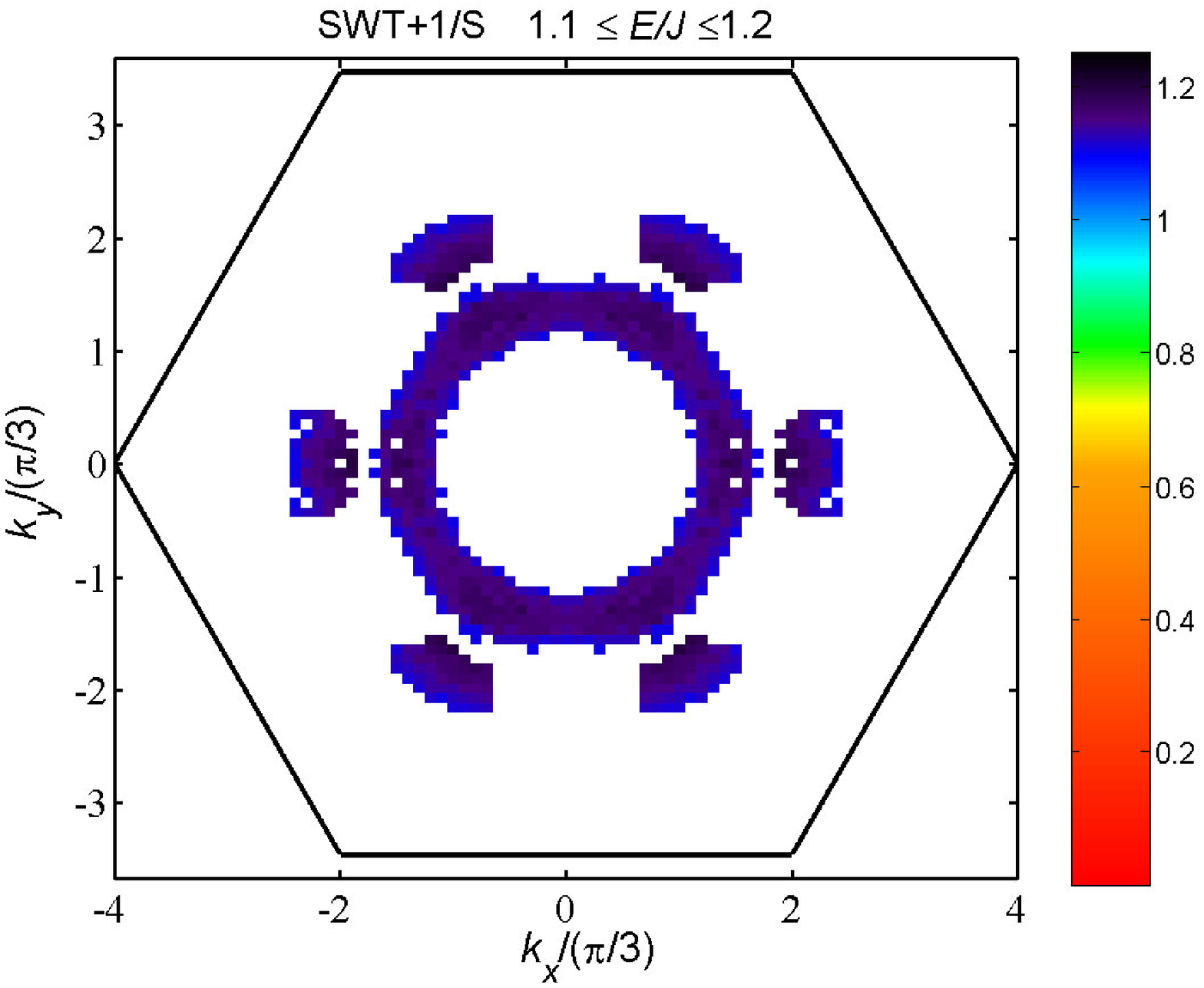}
\includegraphics[scale=0.35]{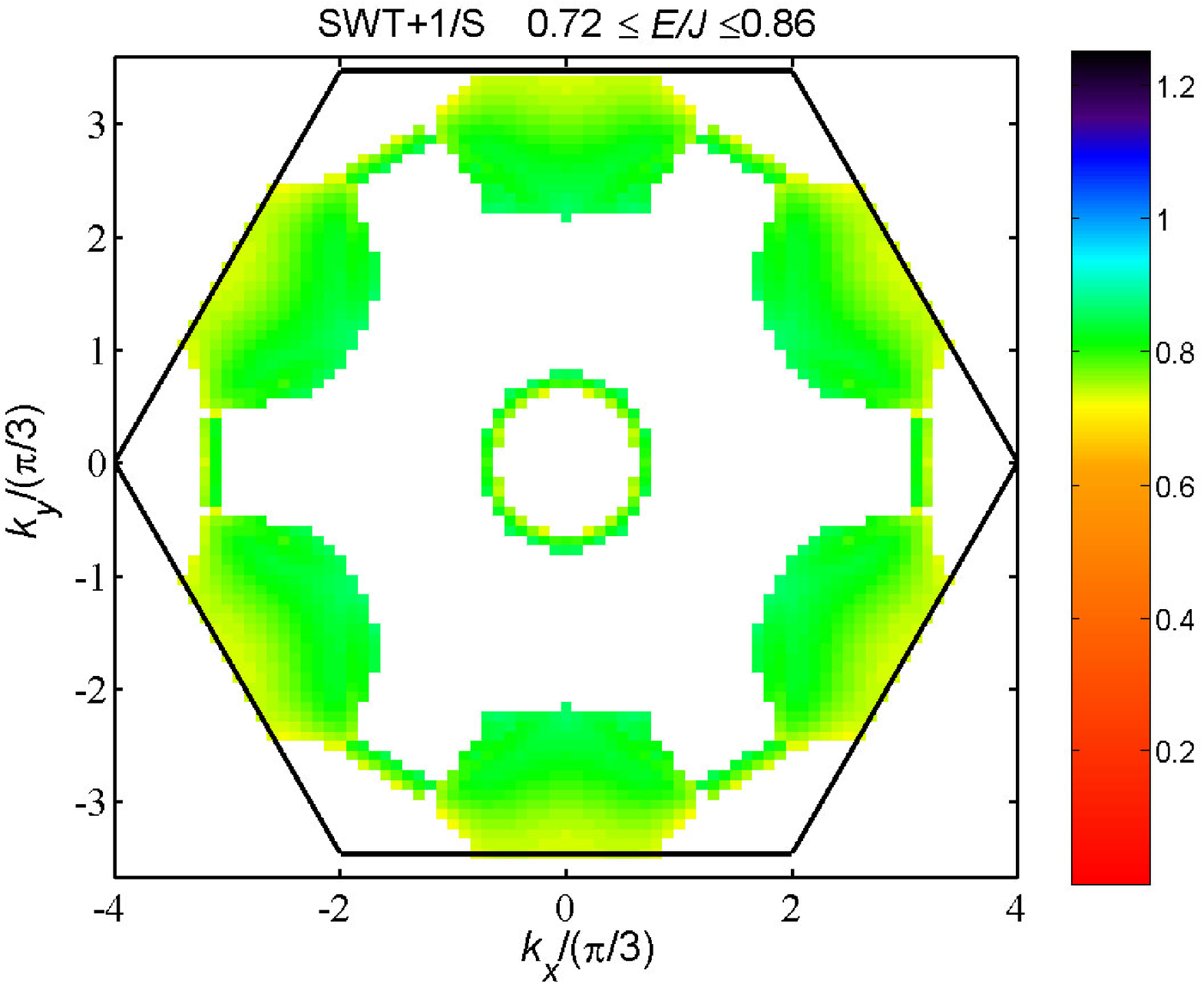}
\caption{\label{dos-peak-swt} (Color online) Highlight of regions
in the Brillouin zone that contribute to the DOS peaks
in the SWT+1/S calculations.}
\end{center}
\end{figure}

4. Both the series results and SWT+1/S results show relatively flat
or dispersionless spectra over large parts of the Brillouin Zone.
These lead to sharp peaks in the density of states. However, there
are some qualitative and quantitative differences here.
In the series results the flattest part of the spectra that gives rise
to the largest peak in the DOS are near the highest energy.
A second smaller peak in the DOS primarily gets contributions from
the neighborhood of the roton minima. Both these regions are highlighted in
Fig.~\ref{dos-peak-series}.
In contrast, in SWT+1/S results the peak in the DOS near the highest magnon
energies is much smaller. The flattest part of the spectra in SWT+1/S
calculations are from the region near the roton minima. These are highlighted
in Fig.~\ref{dos-peak-swt}.

5. The roton minima at wavevector B and equivalent points are present
in both series expansion results and in SWT+1/S results. However, they
are much more pronounced in the series results. A similar roton minima
is seen in the square-lattice at ${\bf k}=(\pi,0)$, where it is also
more prominent in series expansion and quantum Monte Carlo
results,\cite{singh95,sylju1,sandvik3}
absent in SWT+1/S results and barely visible in the next higher
order spin-wave results.\cite{igarashi05}

6. The two-dimensional plots for both series expansions and SWT+1/S have
a similar look with a central annular high energy region, which is separated
from 6 high energy lobes by a minima in the middle. The annular region in
SWT+1/S appears more circular than in the series results, although both
have clear hexagonal features. The lobes also have some differences.

7. The SWT+1/S calculations also predict finite lifetimes for spin-waves
living around the center of the Brillouin Zone. These have not been
taken into account in the series calculations and may also contribute
to the difference between the two spectra.

Overall the comparison shows that the SWT+1/S results have many common
features but also some differences.
Neutron scattering spectra on a triangular-lattice material would
be very exciting to compare with. In the meanwhile, sharp peaks in DOS
may already be singled out in optical measurements.
Given the qualitative differences, such measurements
should be able to differentiate bwteen the SWT+1/S and series results.
However, such spectra would depend on various matrix elements, and
it would be important to develop a detailed theory for Raman scattering
for these systems.

\section{Finite temperature properties}
\label{thermo}

In this section we discuss the implications of the spectra we
have calculated, particularly the rotons, to
properties of the triangular lattice model at finite temperatures.
To emphasize the importance of this issue we first discuss
the renormalized classical behavior expected at low temperatures
for two-dimensional quantum spin systems,
and the results of earlier high temperature series expansions
which suggested otherwise.

\subsection{Finite-temperature anomalies}
\label{finiteT}

For a two-dimensional quantum spin system with an
ordered ground state, the low temperature behavior
should correspond to a Renormalized Classical (RC) one, that is
a ``classical'' state with interacting Goldstone modes
which is captured by the non-linear sigma model.\cite{chn89}
It is instructive to compare the behavior of square
and triangular lattices.
Spin wave theory suggests that there are not
significant differences between the quantum corrections for square and triangular
lattice models at zero temperature. For both
lattices, a diverse range of theoretical calculations suggest
that the reductions in sublattice magnetization and spin-stiffness are comparable.
If this is the case then one might expect Renormalized Classical
behavior in the model to hold upto comparable temperatures.
The relevant model for the square lattice is the O(3) model,
and it has been very successful at describing both experimental
results and the results of numerical calculations on the lattice model.\cite{chn89}

For the triangular lattice, there are three Goldstone modes, two with
velocity $c_\perp$ and one with velocity $c_\parallel$.
The corresponding spin stiffnesses are denoted,
 $\rho_\perp$ and $\rho_\parallel$.
Several different models have been suggested to
be relevant, including O(3)xO(3)/O(2)\cite{nlsmtr} and SU(2).\cite{css}
All of these models  predict similar temperature
dependences for many quantities.

In the large $N$ expansion, including fluctuations
to order $1/N$ (the physical model has $N=2$),
the static structure factor at the ordering wavevector is\cite{css2}
\begin{equation}
S(Q) \simeq 0.85
\left({T \over 4 \pi \rho_s } \right)^4
\xi(T)^2
\label{eqn:sq}
\end{equation}
where the correlation length $\xi(T)$ (in units of the
lattice constrant) is
given by\cite{caution}
\begin{equation}
\xi(T) = 0.021 \left(c \over \rho_s \right)
\left({4 \pi \rho_s \over T} \right)^{1/2}
\exp \left({4 \pi \rho_s \over T} \right)
\label{eqn:xi}
\end{equation}
where $c= (2c_\perp +  c_\parallel)/3$
and $\rho_s=(2\rho_\perp +  \rho_\parallel)/3$
 is the zero-temperature spin stiffness, which sets the temperature
scale for the correlations.
These expressions are quite similar to those for
the $O(3)$ model that is relevant to the square lattice,
with the $4\pi$ replaced by $2\pi$.
An important prediction of equations
(\ref{eqn:sq}) and (\ref{eqn:xi})
is that a plot of
$ T \ln (S(Q))$ or $T \ln (T\xi^2(T))$
 versus temperature
at low temperatures should increase with decreasing temperature and
converge to a finite non-zero value which is proportional to the
spin stiffness in the ordered state at zero temperature. Indeed, the
relevant plots for the spin-1/2 square lattice model\cite{elstner}
and the classical triangular lattice model\cite{southern} do show
the temperature dependence discussed above. However, in contrast,
the plots for the spin-1/2 model on the triangular lattice do not.
In particular, $ T \ln (S(Q))$ or $T \ln (T\xi^2(T))$
 are actually {\it decreasing} with decreasing
temperature\cite{elstner} down to $0.25J$.
This is what one would expect if the ground state
was actually quantum disordered with a finite correlation
length at zero temperature.
 Hence, to be consistent with the
ordered ground state at zero temperature these quantities must
show an upturn at some much lower temperature.

The zero temperature value of the spin stiffness has been estimated
for the TLM by a
variety of methods, as shown in Table \ref{encomparisons}.
The values are in the range $0.06J$ to $0.09J$.
For $\rho_s = 0.06J$ and $c=Ja$ taken from non-linear
spin wave theory (also consistent with the dispersion
relation found by series expansions),
Equation (\ref{eqn:xi})
implies that the correlation length should be
about 0.6 and 12 lattice constants at  temperatures
of $T=J$ and $T=0.25J$, respectively.
For comparison, the high temperature series expansions give\cite{elstner}
values of about 0.5 and 1.5 lattice constants, at $T=J$ and $T=0.25J$, respectively.
It should be noted that the definitions of the
correlation length in the field theory and in the series
expansions is slightly different.\cite{css2}

Furthermore, the entropy for the non-linear sigma model at low temperatures is
just that of non-interacting bosons in two dimensions,
\begin{equation}
s(T) =  \mathcal{A} \left({ 1 \over c_\parallel^2 }
 +  { 2 \over c_\perp^2} \right) T^2 + O(T^4)
\label{eqn:entropy}
\end{equation}
where $\mathcal{A}$ is a dimensionless constant of O(1). SWT+1/S gives $c_\parallel=1.11J$ and
$c_\perp=0.69J$.\cite{css2,chubukov} This means that for $T \ll  J $ the
system should have very small entropy. Indeed for the square lattice
this is the case: it is about 0.05 at $T=0.3J$. Quantum Monte Carlo calculations
found that for $T<0.25J$ the internal energy for the square lattice had the
corresponding $T^3$ dependence.\cite{troyer}
However, for the triangular lattice
the entropy is still 0.3 at $T=0.3J$.
Chubukov, Sachdev, and Senthil,\cite{css2}
suggested that the origin of the above
discrepencies was related to a crossover between
quantum critical and renormalised
classical regimes. Previously, we suggested\cite{zheng06}
that the above discrepancies could be explained
if one considered the rotons to be composed of
a spinon and anti-spinon which were excited thermally.
However, we now show how thermal excitations of rotons can
explain the above discrepencies.
There is a significant analogy  here with the
role that rotons play in superfluid $^4$He where they
start to make substantial
contributions to the entropy at temperatures much
less than the roton gap.\cite{thanks,fig-roton}

\subsection{Contributions of rotons to finite temperature
properties}

\begin{figure}[h]
\begin{center}
  \includegraphics[width=6cm,angle=270]{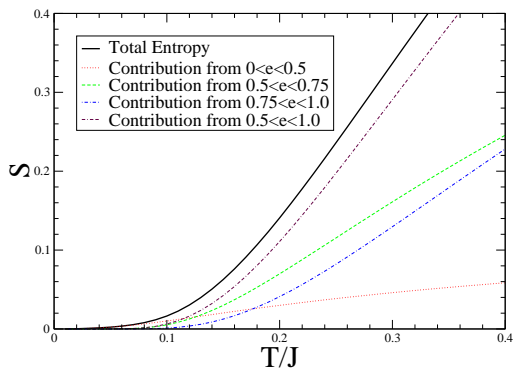}
  \caption{\label{entropy}Entropy of the triangular-lattice Heisenberg model
due to the magnon excitations. Contributions to the entropy from
different energy ranges are shown. The low energy magnons only dominate
the entropy below $T=0.1J$. }
\end{center}
\end{figure}

We will calculate the entropy of the triangular-lattice model
by assuming that the magnon excitations can be treated as a gas of noninteracting
bosons with a dispersion as obtained from the series calculations. The entropy
per site for noninteracting bosons (measured in units of $k_B=1$) is given by
\be
s(T) =\int_0^{\infty}d\varepsilon\;g(\varepsilon)\;\Bigg[\frac{\varepsilon/T}{e^{\beta
\varepsilon}-1} -\ln\Big(1-e^{-\beta\varepsilon}\Big)\Bigg]
\ee
where the DOS $g(\varepsilon)$ is normalized to unity. A plot of this entropy calculated
from the series DOS in Fig. \ref{dosplot} is shown in Fig. \ref{entropy}. It is seen that
the entropy is in fact larger than 0.3 at $T=0.3J$ and thus consistent with the
high-temperature series data. It is also clear that the contribution
to the entropy from the rotons/high-energy excitations (energy $\simeq 0.5$ and above) starts
dominating over the contribution from the low-energy Goldstone modes at a temperatures slightly
above $T=0.1 J$ which is only about 1/5 of the roton gap. This shows that for the triangular
lattice model the presence of rotons significantly influences thermodynamic properties even at very low
temperatures and thus provides an explanation of why the finite-temperature behavior of the
triangular-lattice model are very different from the square lattice model. In particular it implies
that a nonlinear sigma-model description will be valid only at very low temperatures.

\section{Discussion}
\label{disc}

We have seen that SWT+1/S results for the spectra share many features of the series
expansion results. Furthermore, the existence of
rotons and flat-regions in the spectra at fairly low energies
(about 4-times lower than for the square-lattice), gives a natural explanation for why
the results of high-temperature series expansions
for the square and triangular lattice models are
qualitatively different.
However, there are still some important questions to be addressed.
First, we should stress that we still lack a physical picture
(such as exists for superfluid $^4$He, thanks
to Feynman\cite{feynman}) of the nature of the rotons.
Moreover, recent experimental results on
$\kappa$-(BEDT-TTF)$_2$Cu$_2$(CN)$_3$,
and recent variational RVB calculations with spinon excitations,
raise further issues.
We now briefly review these points.

\subsection{Are magnons  bound spinon-anti-spinon pairs?}

In our earlier paper\cite{zheng06} we suggested a possible
explanation of the `roton' minima in the magnon dispersion relation
in terms of a downward energy renormalization due to a
level-repulsion from a higher-energy two-spinon continuum. This
requires that the spinon dispersion have local minima at specific
wave vectors. For the square-lattice model a similar interpretation,
based on the $\pi$-flux phase,\cite{affmar} was originally proposed
by Hsu\cite{hsu} (see also Ref. \onlinecite{hoetal}) to explain the
minima observed at ($\pi,0$) in that case. For the
triangular-lattice model, several RVB states have spinon excitations
with minima at the required locations. For example, Lee and
Feng\cite{leefeng} and Ogata\cite{ogata} considered a Gutzwiller
projected BCS state with $d_{x^2-y^2} + i d_{xy}$ pairing symmetry
in the Gutzwiller approximation. However, the variational energy of
this state ($-0.484J$) is about 15 per cent higher than the best
estimates of the true ground state energy (see Table
\ref{encomparisons}). In contrast, the Gutzwiller projected BCS
states recently studied by by Yunoki and Sorella\cite{yunsor} using
variational Monte Carlo have energies comparable to the best
estimates of the ground state energy. In their study, a state which
can be related to a short range RVB state, has very good variational
energy, and has spinons with mean field dispersion relation
\begin{equation}
E(\bm{k}) = \left(\mu^2 + (\Delta)^2( \cos^2k_1  + \cos^2k_2  +
\sin^2(k_1 -k_2)
  ) \right)^{1/2}
\label{eqn:spinon}
\end{equation}
where $k_1$ and $k_2$ are the components of $\bm{k}$ that are
parallel to the reciprocal lattice vectors, $\bm{G}_1$ and
$\bm{G}_2$, of the triangular lattice. This dispersion has local
minima at the four wave vectors, $\bm{k} = {1 \over 4} \left(\pm
\bm{G}_1 \pm \bm{G}_2 \right).$
 Spinon-antispinon excitations which make up spin triplet
excitations will then have local minima at the six
points in the middle of the edges of the Brillouin zone,
i.e., the location of the roton minima found in the series expansions.

\subsection{Experimental results}
\label{materials}

We now review recent experimental results on
the Mott insulating phase of
$\kappa$-(BEDT-TTF)$_2$Cu$_2$(CN)$_3$.
Very interestingly, this material does not show any magnetic
long-range order down to the lowest temperature studied, 32
mK,\cite{shimizu} despite the fact that this temperature is four
orders of magnitude smaller than the exchange interactions estimated
to be around 250 K.

The temperature dependence of the Knight shift and nuclear magnetic
relaxation rate, $1/T_1$, associated with $^{13}$C nuclei which have a significant
interaction with the electron spin density
have also been measured for this material.\cite{kawamoto,shimizu2}
This is a particularly useful measurement because the
relaxation rate gives a measure of the range of
the antiferromagnetic correlations.
The observed temperature dependence of the Knight shift is the same as that
of the uniform magnetic susceptibility,\cite{shimizu} as it should be.
As the temperature decreases, the ratio $1/T_1 T$ increases by a factor
of about two from 300 K down to 10 K, at which it decreases
by about thirty per cent down to 6 K.
There is no sign of splitting of NMR spectral lines, as would
be expected if long range order develops.
In contrast, for $\kappa$-(BEDT-TTF)$_2$Cu$_2$[N(CN)$_2$]Cl material,
$1/T_1T$ increases rapidly with decreasing temperature
and exhibits a cusp at the Neel temperature, reflecting the diverging
antiferromagnetic correlation length. Evidence for the existence of
magnetic order, in the latter material,
comes from the splitting of NMR lines at low temperatures.\cite{kawamoto0}

The observed temperature dependence of $1/T_1T$ and
the spin echo rate $1/T_2$ for
$\kappa$-(BEDT-TTF)$_2$Cu$_2$(CN)$_3$ is distinctly
different from that predicted by a non-linear sigma model
in the renormalised classical regime\cite{css},
namely that $1/T_1T$ be proportional to $T^{5/2} \xi(T)$,
and $1/T_2$ be proportional to $T^{3} \xi(T)$,
where the correlation length is given by (\ref{eqn:xi}).
In particular, if this material has a magnetically ordered state
at low temperatures, then both
$1/T_1T$ and $1/T_2$
should be increasing rapidly with decreasing temperature,
 not decreasing.
In the quantum critical regime, close to a quantum critical
point,\cite{css} $1/T_1 \sim T^\eta$ where $\eta$ is the anomalous
critical exponent associated with the spin-spin correlation
function. Generally, for O(n) sigma models, this exponent is much
less than one. If $\eta > 1$, as occurs for field theories
with deconfined
spinons,\cite{css,alicea} then $1/T_1 T$ decreases with decreasing
temperature, opposite to what occurs when the spinons are confined,
because then $\eta \ll 1$.\cite{chn89}
 It is very interesting that at low
temperatures, from 1 K down to 20 mK, it was found\cite{shimizu2}
that $1/T_1 \sim T^{3/2}$ and $1/T_2 \sim $ constant.
 In contrast for the materials described
by the Heisenberg model on a square lattice\cite{sandvik2}
 or a chain,\cite{sandvik}
both
relaxation rates {\it diverge} as the temperature decreases. Hence,
these NMR results are clearly inconsistent with a description of the
excitations of this material in terms of interacting magnons. It is
also observed that a magnetic field induces  spatially non-uniform
local moments.\cite{shimizu2} Motrunich has given a
 spin liquid interpretation of this observation.\cite{motrunich2}
The simplest possible explanation of why these results at such
low temperatures are inconsistent with what one
expects for the
nearest neighbour Heisenberg model is that
such a model may not be adequate
to describe this material and the spin liquid state
may arise from the presence of ring exchange terms in
the Hamiltonian.\cite{zheng05hight}
This has led to theoretical studies of such models.\cite{motrunich,leelee}

\section{Conclusions}
\label{concl}

The present comparisons of the series results of the spectra with
the order $1/S$ spin-wave theory suggests that the shape of the
magnon dispersion relation can be understood in a
more conventional picture of interacting magnons.
Furthermore, the existence of the roton minima
at points in the middle of the edges of the Brillouin zone
and regions of flat dispersion in the zone,
can explain why the low temperature properties
of the triangular lattice model are so
different from those of the square lattice.
However, we still lack a clear physical picture
for the nature of the rotons.
An important issue to resolve is whether the
most natural description for them is in terms
of bound spinon-antispinon pairs.

From a theoretical point of view it may be interesting to add other
destabilizing terms to the Hamiltonian, such as second neighbor
interactions and ring-exchange terms, which can demonstrably lead to
short correlation lengths and destabilize the 120 degree order, and
then explore the changes in the dispersion relation, one-magnon
weight and continuum lineshapes with variations in the model
parameters.

It is also important to examine these results in the context of
the organic material $\kappa$-(BEDT-TTF)$_2$Cu$_2$(CN)$_3$.
The measured uniform susceptibility for this material shows excellent
parameter-free agreement with the calculated susceptibility for
the spin-1/2 triangular-lattice Heisenberg model. Yet, this
system does not develop long-range order down to $T/J\approx
10^{-4}$.

\begin{acknowledgments}

We especially thank A. V. Chubukov  and O. A. Starykh
for very helpful discussions and for
sharing their unpublished results with us.
 We also acknowledge helpful
discussions with G. Aeppli, J. Alicea, F. Becca, M.~P.~A. Fisher,
S.~M. Hayden, J. B. Marston,
D.~F. McMorrow, B. J. Powell, S. Sorella, and E. Yusuf. This work
was supported by the Australian Research Council (WZ, JOF, and RHM),
the US National Science Foundation, Grant No. DMR-0240918 (RRPS),
and the United Kingdom Engineering and Physical Sciences Research
Council, Grant No. GR/R76714/02 (RC). We are grateful for the
computing resources provided by the Australian Partnership for
Advanced Computing (APAC) National Facility and by the Australian
Centre for Advanced Computing and Communications (AC3).

\end{acknowledgments}

\newpage
\widetext



\begin{table}
\caption{Series coefficients for the ground state energy per site $E_0/N$ and the order parameter $M$ for
$t=0$ and $t=1$ for the isotropic triangular-lattice model ($q=2\pi/3$).
Series coefficients of $\lambda^n$ up to order $n=13$ are listed.}
\label{tab_e0_M_ser}
\begin{tabular}{|l|l|l|l|l|}
\hline \hline
\multicolumn{1}{|c}{$n$} & \multicolumn{1}{|c}{$E_0/N$ for $t=0$} & \multicolumn{1}{|c}{$E_0/N$ for $t=1$} &
\multicolumn{1}{|c}{$M$ for $t=0$} & \multicolumn{1}{|c|}{$M$ for $t=1$} \\
\hline
  0 &  -3.750000000$\times 10^{-1}$ &  -3.750000000$\times 10^{-1}$ &  ~5.000000000$\times 10^{-1}$ &  ~5.000000000$\times 10^{-1}$ \\
  1 &  ~0.000000000                 &  ~0.000000000                 &  ~0.000000000                 &  ~0.000000000       \\
  2 &  -1.687500000$\times 10^{-1}$ &  -9.375000000$\times 10^{-2}$ &  -1.350000000$\times 10^{-1}$ &  -4.166666667$\times 10^{-2}$ \\
  3 &  ~3.375000000$\times 10^{-2}$ &  -3.125000000$\times 10^{-2}$ &  ~5.400000000$\times 10^{-2}$ &  -2.777777778$\times 10^{-2}$ \\
  4 &  -4.433705357$\times 10^{-2}$ &  -1.435119721$\times 10^{-2}$ &  -1.363457908$\times 10^{-1}$ &  -2.036471287$\times 10^{-2}$ \\
  5 &  ~2.042585300$\times 10^{-2}$ &  -9.090555800$\times 10^{-3}$ &  ~8.589755026$\times 10^{-2}$ &  -1.803575334$\times 10^{-2}$ \\
  6 &  -2.832908602$\times 10^{-2}$ &  -6.546903212$\times 10^{-3}$ &  -1.657631567$\times 10^{-1}$ &  -1.659741503$\times 10^{-2}$ \\
  7 &  ~3.153484699$\times 10^{-2}$ &  -4.684496998$\times 10^{-3}$ &  ~2.055368406$\times 10^{-1}$ &  -1.456148506$\times 10^{-2}$ \\
  8 &  -4.765982794$\times 10^{-2}$ &  -3.395880980$\times 10^{-3}$ &  -3.691101414$\times 10^{-1}$ &  -1.262324583$\times 10^{-2}$ \\
  9 &  ~6.850871690$\times 10^{-2}$ &  -2.535518092$\times 10^{-3}$ &  ~5.890651357$\times 10^{-1}$ &  -1.102173131$\times 10^{-2}$ \\
 10 &  -1.025445984$\times 10^{-1}$ &  -1.940417545$\times 10^{-3}$ &  -1.005494430                 &  -9.680160168$\times 10^{-3}$ \\
 11 &  ~1.565521577$\times 10^{-1}$ &  -1.501987905$\times 10^{-3}$ &  ~1.700641966                 &  -8.486511451$\times 10^{-3}$ \\
 12 &  -2.455267547$\times 10^{-1}$ &  -1.170051241$\times 10^{-3}$ &  -2.948749946                 &  -7.416819496$\times 10^{-3}$ \\
 13 &  ~3.935047914$\times 10^{-1}$ &  -9.185872231$\times 10^{-4}$ &  ~5.156611906                 &  -6.481274769$\times 10^{-3}$ \\
\hline \hline
\end{tabular}
\end{table}


\begin{table}
\caption{Series coefficients for the magnon dispersion for the isotropic triangular-lattice model, calculated
for $t=1$ in Eqs. (\ref{H0}) and (\ref{V}). Nonzero coefficients $c_{r,m,n}$ in Eq. (\ref{eq_mk_y1}) up to order
$r=9$ are listed.}
\label{tab_mk_y1_t1}
\begin{tabular}{|ll|ll|ll|ll|}
\hline \hline \multicolumn{1}{|c}{($r,m,n$)}
&\multicolumn{1}{c|}{$c_{r,m,n}$} &\multicolumn{1}{c}{($r,m,n$)}
&\multicolumn{1}{c|}{$c_{r,m,n}$} &\multicolumn{1}{c}{($r,m,n$)}
&\multicolumn{1}{c|}{$c_{r,m,n}$}
&\multicolumn{1}{c}{($r,m,n$)} &\multicolumn{1}{c|}{$c_{r,m,n}$} \\
\hline
 ( 0, 0, 0) & ~2.500000000     &( 7, 4, 0) & -2.092658337$\times 10^{-2}$ &( 7, 8, 0) & -2.788171252$\times 10^{-3}$ &( 8,12, 0) & ~1.247798114$\times 10^{-4}$ \\
 ( 1, 0, 0) & -1.000000000     &( 8, 4, 0) & -8.143627454$\times 10^{-2}$ &( 8, 8, 0) & -3.352709547$\times 10^{-3}$ &( 9,12, 0) & ~2.470018460$\times 10^{-4}$ \\
 ( 2, 0, 0) & -4.988839286$\times 10^{-1}$ &( 9, 4, 0) & -1.523543756$\times 10^{-2}$ &( 9, 8, 0) & ~7.333948649$\times 10^{-3}$ &( 7,12, 2) & ~2.937003840$\times 10^{-4}$ \\
 ( 3, 0, 0) & -2.740918633$\times 10^{-1}$ &( 3, 5, 1) & ~6.201171875$\times 10^{-2}$ &( 5, 9, 1) & ~2.455267719$\times 10^{-3}$ &( 8,12, 2) & ~1.899234898$\times 10^{-5}$ \\
 ( 4, 0, 0) & -1.128855593$\times 10^{-2}$ &( 4, 5, 1) & -4.072501106$\times 10^{-3}$ &( 6, 9, 1) & -4.769602124$\times 10^{-4}$ &( 9,12, 2) & -6.147374845$\times 10^{-5}$ \\
 ( 5, 0, 0) & ~4.718452314$\times 10^{-2}$ &( 5, 5, 1) & ~1.819088473$\times 10^{-2}$ &( 7, 9, 1) & -1.354096601$\times 10^{-3}$ &( 7,11, 3) & ~4.895006400$\times 10^{-4}$ \\
 ( 6, 0, 0) & ~1.069731871$\times 10^{-2}$ &( 6, 5, 1) & ~3.981234691$\times 10^{-2}$ &( 8, 9, 1) & -5.219152848$\times 10^{-3}$ &( 8,11, 3) & ~3.504866827$\times 10^{-4}$ \\
 ( 7, 0, 0) & -2.783715438$\times 10^{-4}$ &( 7, 5, 1) & -8.391751084$\times 10^{-3}$ &( 9, 9, 1) & -9.505943368$\times 10^{-3}$ &( 9,11, 3) & ~3.548457842$\times 10^{-4}$ \\
 ( 8, 0, 0) & ~1.029634963$\times 10^{-2}$ &( 8, 5, 1) & -4.326852838$\times 10^{-2}$ &( 5, 8, 2) & ~4.910535438$\times 10^{-3}$ &( 7,13, 1) & ~9.790012800$\times 10^{-5}$ \\
 ( 9, 0, 0) & -7.409203681$\times 10^{-3}$ &( 9, 5, 1) & ~1.194051895$\times 10^{-2}$ &( 6, 8, 2) & ~2.089776036$\times 10^{-3}$ &( 8,13, 1) & -1.697239410$\times 10^{-4}$ \\
 ( 1, 2, 0) & ~7.500000000$\times 10^{-1}$ &( 3, 6, 0) & ~1.033528646$\times 10^{-2}$ &( 7, 8, 2) & -1.363667878$\times 10^{-3}$ &( 9,13, 1) & -2.395827580$\times 10^{-4}$ \\
 ( 2, 2, 0) & ~8.035714286$\times 10^{-2}$ &( 4, 6, 0) & -2.807172967$\times 10^{-2}$ &( 8, 8, 2) & -1.944215793$\times 10^{-3}$ &( 7,14, 0) & ~6.992866286$\times 10^{-6}$ \\
 ( 3, 2, 0) & -2.913527716$\times 10^{-1}$ &( 5, 6, 0) & -4.880096703$\times 10^{-2}$ &( 9, 8, 2) & ~2.068274612$\times 10^{-3}$ &( 8,14, 0) & -5.264758087$\times 10^{-5}$ \\
 ( 4, 2, 0) & -3.538253764$\times 10^{-1}$ &( 6, 6, 0) & -3.117757381$\times 10^{-2}$ &( 5,10, 0) & ~2.455267719$\times 10^{-4}$ &( 9,14, 0) & -2.888908886$\times 10^{-5}$ \\
 ( 5, 2, 0) & -1.433040888$\times 10^{-1}$ &( 7, 6, 0) & ~1.625826764$\times 10^{-2}$ &( 6,10, 0) & -7.192654350$\times 10^{-4}$ &( 8,13, 3) & -1.415213825$\times 10^{-4}$ \\
 ( 6, 2, 0) & ~6.032017508$\times 10^{-2}$ &( 8, 6, 0) & ~2.418896967$\times 10^{-2}$ &( 7,10, 0) & ~6.619708053$\times 10^{-5}$ &( 9,13, 3) & -1.012838318$\times 10^{-4}$ \\
 ( 7, 2, 0) & ~5.451813695$\times 10^{-2}$ &( 9, 6, 0) & -3.062453544$\times 10^{-2}$ &( 8,10, 0) & ~6.397714641$\times 10^{-4}$ &( 8,14, 2) & -7.076069125$\times 10^{-5}$ \\
 ( 8, 2, 0) & -4.427446338$\times 10^{-2}$ &( 4, 7, 1) & -1.581420898$\times 10^{-2}$ &( 9,10, 0) & -1.290395255$\times 10^{-3}$ &( 9,14, 2) & ~1.239259101$\times 10^{-5}$ \\
 ( 9, 2, 0) & -5.261673101$\times 10^{-2}$ &( 5, 7, 1) & -1.015665919$\times 10^{-2}$ &( 6,10, 2) & -1.311427742$\times 10^{-3}$ &( 8,12, 4) & -8.845086406$\times 10^{-5}$ \\
 ( 2, 3, 1) & -4.218750000$\times 10^{-1}$ &( 6, 7, 1) & -1.671867122$\times 10^{-2}$ &( 7,10, 2) & -3.880276886$\times 10^{-4}$ &( 9,12, 4) & -8.553417963$\times 10^{-5}$ \\
 ( 3, 3, 1) & ~1.425980548$\times 10^{-1}$ &( 7, 7, 1) & -1.561157615$\times 10^{-2}$ &( 8,10, 2) & ~4.062057862$\times 10^{-4}$ &( 8,15, 1) & -2.021734036$\times 10^{-5}$ \\
 ( 4, 3, 1) & ~2.239011724$\times 10^{-1}$ &( 8, 7, 1) & ~1.139858035$\times 10^{-2}$ &( 9,10, 2) & -1.607660184$\times 10^{-4}$ &( 9,15, 1) & ~4.795929060$\times 10^{-5}$ \\
 ( 5, 3, 1) & -8.857143618$\times 10^{-3}$ &( 9, 7, 1) & ~3.929130986$\times 10^{-2}$ &( 6,11, 1) & -5.245710967$\times 10^{-4}$ &( 8,16, 0) & -1.263583772$\times 10^{-6}$ \\
 ( 6, 3, 1) & -1.556426647$\times 10^{-1}$ &( 4, 6, 2) & -1.186065674$\times 10^{-2}$ &( 7,11, 1) & ~3.733048771$\times 10^{-4}$ &( 9,16, 0) & ~1.192482027$\times 10^{-5}$ \\
 ( 7, 3, 1) & -4.175359076$\times 10^{-2}$ &( 5, 6, 2) & -1.432147845$\times 10^{-2}$ &( 8,11, 1) & ~1.091608982$\times 10^{-3}$ &( 9,15, 3) & ~3.763753619$\times 10^{-5}$ \\
 ( 8, 3, 1) & ~1.015643471$\times 10^{-1}$ &( 6, 6, 2) & -1.435136738$\times 10^{-2}$ &( 9,11, 1) & ~1.851573788$\times 10^{-3}$ &( 9,14, 4) & ~5.645630428$\times 10^{-5}$ \\
 ( 9, 3, 1) & ~4.756044803$\times 10^{-2}$ &( 7, 6, 2) & -5.307444139$\times 10^{-3}$ &( 6, 9, 3) & -8.742851612$\times 10^{-4}$ &( 9,16, 2) & ~1.613037265$\times 10^{-5}$ \\
 ( 2, 4, 0) & -2.109375000$\times 10^{-1}$ &( 8, 6, 2) & -6.778541366$\times 10^{-3}$ &( 7, 9, 3) & -4.761605554$\times 10^{-4}$ &( 9,17, 1) & ~4.032593163$\times 10^{-6}$ \\
 ( 3, 4, 0) & -1.089232568$\times 10^{-1}$ &( 9, 6, 2) & -2.509466719$\times 10^{-2}$ &( 8, 9, 3) & -3.368803230$\times 10^{-4}$ &( 9,18, 0) & ~2.240329535$\times 10^{-7}$ \\
 ( 4, 4, 0) & ~8.339453634$\times 10^{-3}$ &( 4, 8, 0) & -1.976776123$\times 10^{-3}$ &( 9, 9, 3) & -8.706110790$\times 10^{-4}$ &           &                    \\
 ( 5, 4, 0) & ~1.053325085$\times 10^{-1}$ &( 5, 8, 0) & ~3.310043485$\times 10^{-3}$ &( 6,12, 0) & -4.371425806$\times 10^{-5}$ &           &                    \\
 ( 6, 4, 0) & ~9.023634592$\times 10^{-2}$ &( 6, 8, 0) & ~2.627504861$\times 10^{-3}$ &( 7,12, 0) & ~1.915360692$\times 10^{-4}$ &           &                    \\
\hline \hline
\end{tabular}
\end{table}

\begin{table}
\begin{center}
\caption{Ground state properties of the isotropic triangular-lattice
model, obtained by different methods. $E_0/N$ is the energy per site
(in units of $J$) for a system with $N$ lattice sites. The order
parameter $M$ is the value of the expectation value of the spin in
the ordered state. It would have a value of $0.5$ in the absence of
quantum fluctuations, and is zero in a spin liquid ground state.
$\rho_s$ is the average spin stiffness (in units of $J$) which in a
non-linear sigma model description sets the temperature scale of the
finite temperature properties.
Note that some of the spin liquid states
based on variational wave functions
give values for the ground state energy comparable
to the best estimates. DMRG, QMC, V, GF, and ED denote density
matrix renormalisation group, Quantum Monte Carlo, Variational,
Greens function, and
Exact diagonalization, respectively. SRVB denotes short range RVB.
GA denotes the Gutzwiller approximation. SB+1/N denotes
Schwinger boson mean-field theory with 1/N fluctuations.} \vspace{0.5cm}
\label{encomparisons}
\begin{tabular}{|l|l|l|l|l|l|}\hline
Method & Ref. & $N$ & $E_0/N$ & $M$ & $\rho_s$ \\\hline
Series &  this work & $\infty$ & $-0.5502(4)$ & $0.19(2) $&  \\
\hline
ED & \onlinecite{runge,bernu94} & $12$ & $-0.6103$ &  & \\
& & $36$ & $-0.5604$  & 0.40 & \\
\hline
 V SRVB& \onlinecite{sindzingre} & $12$ & $-0.6096$  & 0 & 0 \\
&                                  & $36$ & $-0.5579$  & 0 & 0 \\
\hline
ED & \onlinecite{lecheminant} & 36 &   &  & 0.06 \\
\hline
DMRG& \onlinecite{xiang} & $\infty$ & $-0.5442$ & & \\
\hline
GFQMC & \onlinecite{cap99}  & $\infty$ & $ -0.5458(1)$ & $0.205(10)$&  \\
\hline
VQMC, SRVB & \onlinecite{yunsor}  & $\infty$ & $ -0.5123 $ & $0$& 0 \\
\hline
VQMC, RVB & \onlinecite{yunsor}  & $\infty$ & $ -0.5357 $ & $0$& 0 \\
\hline
SWT+1/S & \onlinecite{miy} & $\infty$  & $-0.5466$  & 0.2497&   \\
\hline SWT+1/S & \onlinecite{chubukov} & $\infty$  &   & 0.266
 & 0.087  \\
\hline
d + id RVB GA & \onlinecite{leefeng} & $\infty$  & $-0.484(2)$  & 0 & 0  \\
\hline
Coupled cluster & \onlinecite{kruger} & $\infty$  &   & 0.2134 & $\rho_\parallel= $ 0.056   \\
\hline
SB+1/N & \onlinecite{manuel} & $\infty$  & -0.5533  &  & $\rho_\parallel= $ 0.09   \\
\hline
\end{tabular}
\end{center}
\end{table}


\begin{thebibliography}{99}
\bibitem{caution}
Note that this is based on equations (4.7) and (4.8) in
\onlinecite{css2}, whereas the prefactor in equation (7.4) assumes
the value of the spin stiffness from SWT+1/S.
\end{thebibliography}
\end{document}